\documentstyle[12pt,axodraw]{article}
\begin{document}


\begin{center}
{\large \bf
Two-Loop Bosonic Electroweak Corrections to the Muon Lifetime
and $M_Z$--$M_W$ Interdependence}
\end{center}
\begin{center}
{
A.~Onishchenko%
\footnote{~E-mail: onish@particle.uni-karlsruhe.de}
\footnote{~Supported by DFG-Forschergruppe  {\it ``Quantenfeldtheorie,
               Computeralgebra und\\ Monte-Carlo-Simulation''}
               (contract FOR 264/2-1)}
and
O.~Veretin%
\footnote{~E-mail: veretin@particle.uni-karlsruhe.de}
\footnote{~Supported by BMBF under grant No 05HT9VKB0.}
}

\vspace{5mm}
{
Institut f\"ur Theoretische Teilchenphysik,\\
Universit\"at Karlsruhe, D-76128 Karlsruhe, Deutschland
}
\end{center}
\hspace{3in}
\begin{abstract}
The two-loop electroweak bosonic correction to the muon lifetime
and the $M_W$--$M_Z$ interdependence is computed using analytical
methods of asymptotic expansion. Combined with previous
calculations this completes the full two-loop correction
to $\Delta r$ in the Standard Model. The shift to the prediction
of $W$-boson mass due to two-loop bosonic corrections does not
exceed 1 MeV for the range of the Higgs boson mass from 100 to 1000 GeV.
\end{abstract}

\vspace{10mm}

The Fermi constant $G_F$ plays an important role in the precision tests
of the Standard Model. Theoretically $G_F$ can be related to other
precision observables: the electroweak coupling constant $\alpha$
and the masses of
electroweak gauge bosons $M_Z$ and $M_W$. Other parameters enter
this expression through quantum corrections.
Usually one inverts this
relation in order to predict $M_W$ through $M_Z$ which is measured
much more accurate. This $M_Z-M_W$ interdependence can be then
confronted with experimental value $M_W^{\rm exp}$.
The current error (39 MeV) of $M_W^{\rm exp}$
will be drastically reduced at future colliders.
In fact, at LHC the experimental error can be reduced
to 15 MeV \cite{LHC} and at Linear Collider even down to 6 MeV \cite{TESLA}.
Therefore much efforts have been spent to reduce the error of
the theoretical prediction.

Fermi constant is defined as the coupling constant
in the low energy four fermion effective Lagrangian
describing the decay of the muon
\begin{equation}
\label{lagrangian}
{\cal L}_F = {\cal L}_{\rm QED}
    + \frac{G_F}{\sqrt{2}} \,
    \bigl[\bar{\nu}_\mu \gamma^\alpha (1-\gamma_5) \mu \bigr]
    \,   \bigl[  \bar{e} \gamma_\alpha (1-\gamma_5) \nu_e \bigr]
\end{equation}
where $e$ and $\mu$ are electron and muon fields, $\nu_e$ and $\nu_\mu$
are the corresponding neutrinos and $G_F$ is the Fermi constant.
From the Lagrangian (\ref{lagrangian}) one gets
the following value for the muon lifetime
\begin{equation}
\frac{1}{\tau_\mu} =
\frac{G_F^2 m_\mu^5}{192\pi^3}
   \left( 1 - 8\frac{m_e^2}{m_\mu^2} \right) (1+\Delta q) \,,
\end{equation}
where the factor $\Delta q$ describes all the quantum corrections
in the low energy effective theory (i.e. QED corrections).
At one-loop order
these corrections have been computed a long time ago \cite{QEDoneloop}.
Recently also the two-loop result for $\Delta q$ has been obtained
\cite{QEDtwoloop}. Taking it into account,
the error of $G_F$ is nowadays dominated by the experimental error of
$\tau_\mu$ measurement\footnote{The present value is
$G_F = 1.16637(1)\times 10^{-5} \,\,\mbox{GeV}^{-2}$
and possible future experiments could reduce the error
by an order of magnitude.}.

  In order to relate $G_F$ to $M_W$ one can use
the matching condition between effective theory (\ref{lagrangian})
and the Standard Model, which requires that the value of $\tau_\mu$
does not depend on whether it is evaluated in the Fermi theory
or in the full Standard Model. This brings us to the relation
\begin{equation}
 \frac{G_F}{\sqrt{2}} = \frac{e^2}{8 s_W^2 M_W^2} (1+\Delta r),
\label{GFrelation}
\end{equation}
where $e$ is the electric charge,
$s_W=\sin\theta_W$ is the SM mixing angle and $M_W$ is the $W$-boson mass.
The factor $e^2/8s_W^2m_W^2$ corresponds to the matching at the three
level, while
the quantity $\Delta r=\Delta r^{\rm 1-loop}+\Delta r^{\rm 2-loop}+\dots$
is defined to absorb quantum corrections coming from the matching
procedure at higher loop orders.
At the one-loop level $\Delta r$ was first evaluated in \cite{DRoneloop}.
Further improvement of $\Delta r$ was achieved in \cite{Largetop}
where the leading and subleading
large $t$-quark corrections were computed at $O(\alpha^2)$ order.
Other two-loop fermionic corrections were
considered in a series of papers \cite{Weiglein}.

  The aim of this work is to compute the remaining $O(\alpha^2)$
pure bosonic contribution in order to complete the full
$O(\alpha^2)$ approximation to $\Delta r$. As this work was in
progress Ref. \cite{Czakon} has appeared where this missing
contribution has been computed numerically using  the integral
representation for the massive two-point functions
\cite{Bauberger:1994by}.

  Our approach is complementary to that of \cite{Czakon}.
We use the methods of asymptotic expansion \cite{asymptotic}
in two different regimes: 1) large Higgs mass expansion and 2) expansion
in difference of masses $M_H^2-M_Z^2$. In addition we consider also
$s_W^2=\sin^2\theta_W$ as a small parameter in the spirit of \cite{bosons}.

  The details of our calculation will be presented in a separate
publication \cite{Awramik:2002vu} and here we only
sketch the most important moments.
\begin{itemize}
\item
$G_F$ is nothing but the Wilson coefficient function of the corresponding
operator in the low energy four fermion effective theory. Therefore it
depends only on the ``hard'' physics and is insensitive to low energy
field modes in loops.
The simplest way to compute such a quantity is the use
of the factorization theorem (if it exists), which allows one
to separate ``hard'' and ``soft'' physics%
\footnote{In our case by ``soft'' we understand energies and
momenta of order of muon mass $m_\mu$, while ``hard'' means
scales $\gg m_\mu$}.
This theorem for the construction of low energy effective theories
is known for a long time (see e.g. \cite{Gorishnii}) and is based on
the euclidean structure of large mass expansion.
\item
Using the factorization theorem the problem is reduced to
the evaluation of two-loop bubble diagrams.
\item
All the calculations have been performed in a general $R_\xi$ gauge
with three arbitrary gauge parameters. By explicit calculation
we checked that contribution to $\Delta r$ is gauge invariant.
\item
In order to have explicit gauge invariant expressions also at
the level of bare quantities we explicitely include tadpole diagrams.
We have checked that applying this prescription all
the bare parameters and all
the counterterms for the masses and the coupling
constant are explicitely gauge invariant.
\item
The renormalization was performed
in two different schemes: $\overline{\rm MS}$
and on-shell scheme. For the transformation from one scheme to
another one needs the results of \cite{bosons} where the transformation
formulae for the gauge boson masses were given to two loops.
\end{itemize}
To obtain the two-loop  bosonic contribution to $\triangle r$
a number of two-loop diagrams of the order of $10^4$ should be evaluated.
In our calculation the computer algebra system FORM \cite{FORM} was used.
In order to obtain the FORM readable input we make use of the
input generator DIANA \cite{DIANA}.

%
%
\begin{figure}[h]
\centerline{
\raisebox{30mm}{$\Delta r^{(2)}_{\rm bos}$}\hspace{-8mm}
\raisebox{58mm}{$\times10^{-5}$}\hspace{-4mm}
{\epsfysize=60mm \epsfbox{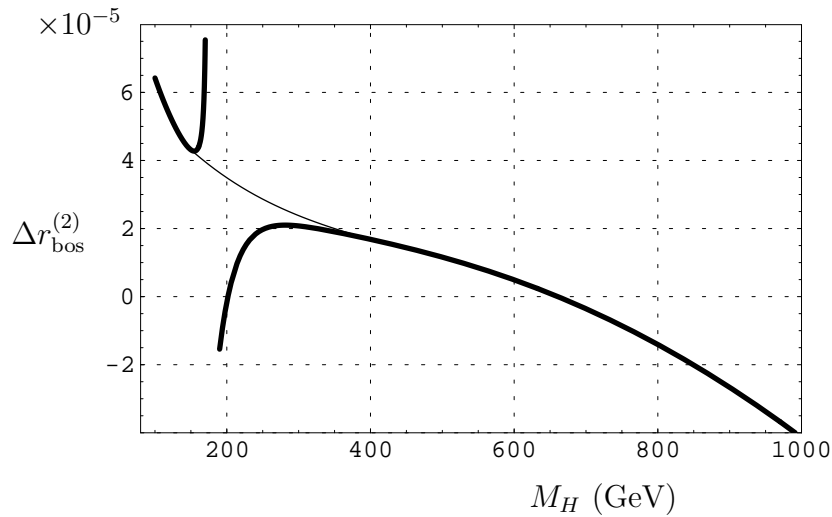}}
\hspace{-43mm} \raisebox{-5mm}{$M_H \,\,(\mbox{GeV})$}
}
\caption{\label{deltar}
  Dependence of $\Delta r^{(2)}_{\rm bos}$ on Higgs mass $M_H$
  (see explanation in the text).
  Parameter choice for the pole masses: $M_Z=91.1876 \,{\rm GeV}$
  and $M_W=80.423 \,{\rm GeV}$.
}
\end{figure}

  Let us now present the results of the calculation.
In Fig. 1 the results of expansions are plotted as a function of
the on-shell Higgs boson mass $M_H$ starting from the experimental
low bound 114 GeV \cite{Hagiwara:pw} up to 1 TeV. First we expand
in the limit $M_H\to M_Z$, e.g. in the difference $(M_H^2-M_Z^2)$.
By summing this series we have found that with 6-10 coefficients
it works well for the values of $M_H$ down to $\sim125$ GeV. In
order to accelerate the convergence of this series we apply P\'ade
approximation w.r.t. the variable $y=(M_H^2-M_Z^2)/M_Z^2$. The
left bold curve in Fig. 1 represents the [3/3] P\'ade approximant
which behaves smooth down to $\sim170$ GeV. The other bold curve
is a sum of the first 4 terms of large Higgs mass expansion. This
series appeares to be nonalternating and the application of P\'ade
did not show any considerable improvement. In the intermediate
region $155\mbox{ GeV}<M_H<350\mbox{ GeV}$ both approximations
fail to converge. (This is due to the presence of singularities in
$M_H$-plane at $|M_H|=2M_W$ or $|M_H|=2M_Z$. Such singularity
appears for example in diagram of Fig. 2. This manifests itself as
a divergence of the series as $M_H$ approaches $2M_Z$.) Therefore,
in order to get a result for Higgs mass values in the region
155--350 GeV we made simple polynomial interpolation.

\begin{center}
\begin{figure}[h]
\centerline{\epsfysize=30mm \epsfbox{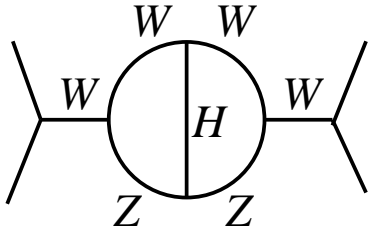}}
\caption{\label{diagram}
  Example of a diagram that has singularity near $M_H=2M_W$.
}
\end{figure}
\end{center}

  Our comparison with the numerical curve from Ref. \cite{Czakon}
shows a perfect agreement between the two calculations. The difference
for  $155\mbox{ GeV}<M_H<350\mbox{ GeV}$ does not exceed 5\% while
for $M_H$ below 155 GeV and above 350 GeV it is much smaller.

\begin{figure}[h]
\centerline{
\raisebox{30mm}{$\Delta M_W^{(2),\rm bos}$}\hspace{-17mm}
\raisebox{24mm}{(MeV)}\hspace{-0mm}
{\epsfysize=60mm \epsfbox{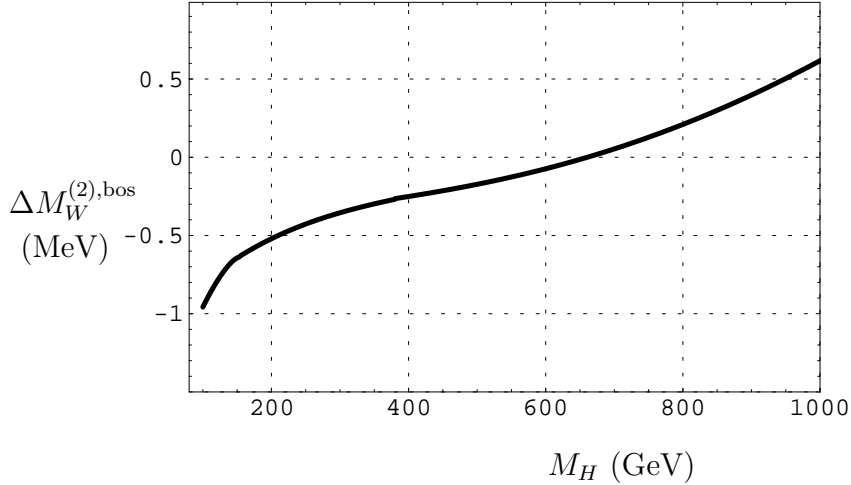}}
\hspace{-43mm} \raisebox{-5mm}{$M_H \,\,(\mbox{GeV})$}
}
\caption{\label{MWmassshift}
  Shift $\Delta M_W$ of the predicted mass of $W$-boson due to two-loop
  bosonic correction to $\Delta r$.
}
\end{figure}

  Let us now turn to the $M_Z-M_W$ interdependence.
By inverting Eq. (\ref{GFrelation}) we can find how
much does the computed $\Delta r^{(2)}_{\rm bos}$ contribute
to the shift of the value of the $W$-boson mass.
This contribution is plotted in Fig. 3.
For the broad region of $M_H$ from 100 to 1000 GeV this correction
does not exceed 1 MeV and appears to be insignificant.
Therefore, we can see that the uncertainty in $M_W$ prediction
at the moment is dominated
by the uncertainty in the experimental value of $t$-quark pole mass.

  In conclusion, the bosonic contributions to $\Delta r$ has
been computed using methods of asymptotic expansions and the low
energy factorization theorem. The influence of this correction on
the theoretical prediction of $M_W$ is found. The gauge invariance
of $\Delta r^{(2)}_{\rm bos}$ has been proved explicitly. By the
inclusion of Higgs-tadpole diagrams it has been checked that all
mass- and charge- counterterms are gauge invariant.

\noindent
{\it Acknowledgments:}
Authors would like to thank M. Kalmykov, M. Awramik, M. Czakon,
K. Chetyrkin and J. K\"uhn for fruitful discussions and
F. Campanario for the careful reading the paper. We are also
grateful to M. Tentyukov for his help with input generator DIANA.




\end{document}